\documentclass[journal=jacsat,manuscript = communication]{achemso}

\usepackage{graphicx}
\usepackage{amssymb}
\usepackage{multirow}
\usepackage{booktabs}
\usepackage{bm}
\SectionNumbersOn
\SectionsOn
\setkeys{acs}{articletitle = true}
\usepackage{titlesec}

\titleformat*{\section}{\small\bfseries}
\titleformat*{\subsection}{\normalsize\bfseries}

\usepackage[normalem]{ulem}
\usepackage[usenames,dvipsnames]{xcolor} 

\title{Superposition of droplet elasticity and volume fraction effects on emulsion dynamics}

\author{Ryan Poling-Skutvik}
\affiliation{Department of Chemical and Biomolecular Engineering, University of Pennsylvania, Philadelphia, Pennsylvania 19104}

\author{Xiaojun Di}
\affiliation{Department of Chemical and Environmental Engineering, Yale University, New Haven, Connecticut 06520}

\author{Chinedum O. Osuji}
\email{cosuji@seas.upenn.edu}
\affiliation{Department of Chemical and Biomolecular Engineering, University of Pennsylvania, Philadelphia, Pennsylvania 19104}

\date{\today}

\begin{document}

\begin{abstract}

The rheological properties of emulsions are of considerable importance in a diverse range of scenarios. Here we describe a superposition of the effects of droplet elasticity and volume fraction on the dynamics of emulsions. The superposition is governed by physical interactions between droplets, and provides a new mechanism for modifying the flow behavior of emulsions, by controlling the elasticity of the dispersed phase. We investigate the properties of suspensions of emulsified wormlike micelles (WLM).  Dense suspensions of the emulsified WLM droplets exhibit thermally responsive properties in which the viscoelastic moduli decrease by an order of magnitude over a temperature range of 0 $^\circ$C to 25 $^\circ$C. Surprisingly, the fragility (i.e. the volume-fraction dependence of the modulus) of the emulsions does not change with temperature. Instead, the emulsion modulus scales as a power-law with volume fraction with a constant exponent across all temperatures even as the droplet properties change from elastic to viscous. Nevertheless, the underlying droplet dynamics depend strongly on temperature. From stress relaxation experiments, we quantify droplet dynamics across the cage breaking time scale below which the droplets are locally caged by neighbors and above which the droplets escape their cages to fully relax. For elastic droplets and high volume fractions, droplets relax less stress through cage rattling and the terminal relaxations are slower than for viscous droplets  and lower volume fractions. The cage rattling and cage breaking dynamics are highly correlated for variations in both temperature and emulsion concentration, suggesting that thermal and volume fraction effects represent independent parameters to control emulsion properties.

\end{abstract}

\section{Introduction}

Multicomponent systems such as foams (gas-in-liquid),\cite{Conn2014, Weaire2008} emulsions (liquid-in-liquid),\cite{Helgeson2016} and colloidal dispersions (solid-in-liquid)\cite{Hunter2012} undergo significant changes in their bulk properties as the concentration of the dispersed phase increases. For systems in which the dispersed phase interacts through (pseudo)-hard-sphere interactions, the change in mechanical properties as well as the underlying microstructure and dynamics can be mapped onto a single controlling parameter --- the volume fraction $\phi$.\cite{Pusey1986} At low volume fractions, these systems typically behave as Newtonian fluids. For intermediate volume fractions, the dynamics of the dispersed phase are slowed and the dispersions exhibit moderate viscoelasticity until the volume fraction increases beyond the glass or jamming transitions.\cite{VanHecke2010,Pellet2016,Vlassopoulos2014} Above these transitions, the dispersed phase is localized by neighboring particles to suppress their long time dynamics. While this mapping represents a beautiful physical simplicity, it also represents a limitation of the physical properties of the system --- to change the droplet dynamics or modulus of a suspension requires a change in volume fraction. This limitation can be overcome, however, by altering how the dispersed droplets interact.

In highly repulsive systems such as charged colloids dispersed in an organic solvent, the particle dynamics can be suppressed at much lower volume fractions than in hard sphere systems due to an increase in effective particle size.\cite{Park2017} Similarly, in strongly attractive  systems, particles form percolating structures, dynamics fully arrest, and the suspension viscoelasticity increases even at low volume fractions.\cite{Lu2008} Chemical attractions and repulsions control the magnitude of the free energy profile between particles, but there are also methods to change the \emph{shape} or steepness of the interaction profile between particles by targeting the \emph{physical} interactions. For example, grafted layers on particles or particles that are themselves crosslinked networks of polymers elastically compress to make the interaction profile less steep than that of hard spheres.\cite{Camargo2010,Camargo2012, Pellet2016} Studies on these compressible, soft particles observe that the volume fraction dependence of particle dynamics (i.e. fragility) is significantly different than that of hard spheres.\cite{Mattsson2009, Gnan2019} Whereas the particle dynamics in hard sphere suspensions are fragile and diverge near the glass and jamming transitions, the dynamics in suspensions of soft and compressible particles are less fragile and smoothly decrease with increasing volume fraction. Physical interaction profiles can be dynamically controlled by making the dispersed or continuous phases from materials that respond to stimuli such as temperature,\cite{Senff1999,Stieger2004,Eberle2012} pH,\cite{Cho2009,Kan2013} and salt concentration.\cite{Azzaroni2005} In the majority of these systems, however, particle softness is convoluted with compressibility so that the effective volume fraction changes concomitant with the interaction profile. Tuning particle softness independent of particle compressibility will isolate the effects of soft physical interactions on the mechanical properties and dynamics of dense emulsions. One potential avenue to generate an system of incompressible yet soft droplets is by emulsifying a complex fluid such as wormlike micelles. This methodology requires orthogonal surfactants in which one surfactant forms the wormlike micelles inside a droplet and the other stabilizes the emulsion droplet interface.

In this Paper, we successfully develop a model system of emulsified droplets of wormlike micelles (WLMs) dispersed into a continuous oil phase where the aqueous emulsion droplets are incompressible but exhibit variable mechanical moduli. The neat WLM solution is thermoresponsive so that its mechanical properties transition from solid-like to liquid-like over a moderate temperature range $ 0 \:^\circ \mathrm{C} \leq T \leq 25 \:^\circ \mathrm{C}$. Dispersing the WLM solution into an oil phase generates an emulsion with thermoresponsive dynamics and mechanical properties. With increasing temperature, the emulsion behaves less elastically as the dispersed droplets become more viscous. In contrast to earlier reports that softer particles form stronger glasses, the volume-fraction dependence of the emulsion mechanical properties does not change with temperature. Instead, the droplet stiffness controls the underlying dynamics of the emulsion. As the droplets become less stiff, their dynamics become less caged and the droplets escape their local cage more quickly, leading to faster terminal relaxations. This coupling between short and long time dynamics of the emulsion droplets persists across all temperatures and volume fractions, suggesting that thermal and concentration effects superpose in this system and represent independent variables by which to tune emulsion properties.

\section{Materials and Methods}
To produce a solution of worm-like micelles (WLM), Ethoquad O/12PG (EQ, AkzoNobel) and sodium salicylate (NaSal, Sigma Alidrich) are added to deionized water at concentrations of 0.72 mol L$^{-1}$ and 0.15 mol L$^{-1}$ respectively. To produce a compatible oil phase, modified polyether-polysiloxane (EM97, Evonik Industries) is added to cyclohexasiloxane (D6, XIAMETER PMX-0246, Dow Corning) at a 1:9 weight ratio. The aqueous WLM phase is dropwise added to the continuous oil phase at high shear rates in a high-speed rotating mixer (20,000 rpm, IKA T18 Ultra Turrax) until the volume fraction reaches $\phi = 0.74$. Incorporated air is removed by brief centrifugation at $\approx3000\times \mathrm{g}$. This procedure generates a stable emulsion of 2 $\mu$m droplets as measured by dynamic light scattering. Emulsions of different volume fractions are produced by diluting this master batch by adding appropriate amounts of the continuous oil phase, ensuring that the droplet size does not vary between samples. All rheological measurements are performed on a TA ARES LS1 rheometer using a 25-mm cone and plate geometry. For oscillatory measurements, experiments are conducted at an oscillatory strain $\gamma = 1$\%. For stress relaxation experiments, samples are pre-sheared for 15 min, allowed to rest for 30 min, and then subjected to an initial step strain of 1\%. This procedure erases previous shear history and results in reproducible measurements. 

\section{Results and Discussion}

\subsection{Worm-like micelle solution}

\begin{figure}[b!]
\centering
\includegraphics[width = 3.25 in]{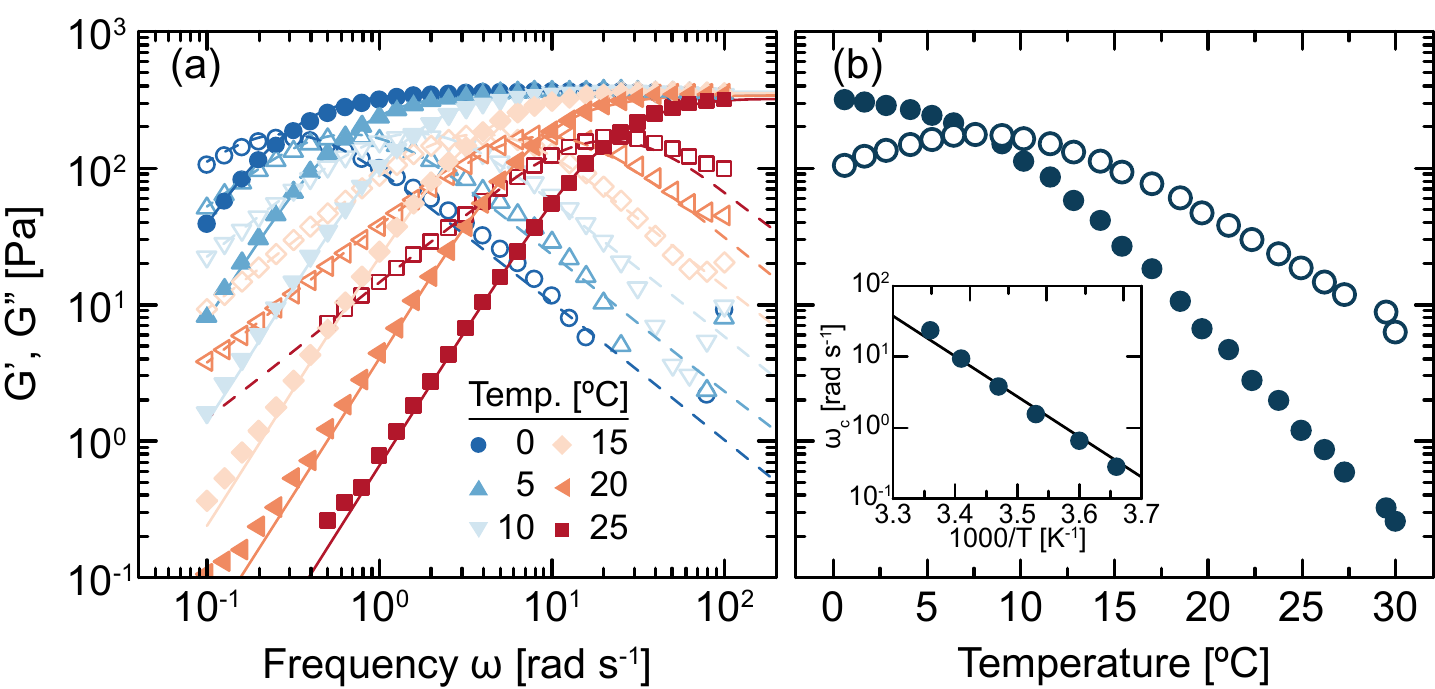}
\caption{\label{fig:WLM_Tdep} (a) Storage $G'$ (closed) and loss $G''$ (open) moduli for the neat WLM solution as (a) a function of frequency $\omega$ for different temperatures and (b) as a function of temperature for $\omega = 1$ rad s$^{-1}$. Solid and dashed curves are fits to a Maxwell model (Eq.\ \ref{eq:Maxwell}) for $G'$ and $G''$, respectively. \textit{Inset:} Temperature dependence of the crossover frequency $\omega_\mathrm{c}$. Solid line is an Arrhenius fit with $|E_\mathrm{a,WLM}| = 120 \pm 20$ kJ mol$^{-1}$.}
\end{figure}

The EQ/NaSal worm-like micelles form a viscoelastic solution with storage $G'$ and loss $G''$ moduli that depend on frequency $\omega$ and temperature (Fig.\ \ref{fig:WLM_Tdep}). These moduli are well-fit by the classical expressions for a Maxwell fluid 
\begin{equation}
\label{eq:Maxwell}
G' = G_\mathrm{e}\frac{\bar{\omega}^2}{1 + \bar{\omega}^2} 
\quad \textrm{and} \quad 
G'' = G_\mathrm{e} \frac{\bar{\omega}}{1+\bar{\omega}^2},
\end{equation}
where $\bar{\omega} = \omega/\omega_\mathrm{c}$ and $\omega_\mathrm{c}$ is the characteristic crossover frequency. While the plateau modulus $G_\mathrm{e}$ does not change with temperature, $\omega_\mathrm{c}$ exhibits an Arrhenius temperature dependence with an activation energy magnitude $|E_\mathrm{a,WLM}|= 120 \pm 20 $ kJ mol$^{-1}$. Due to this shift in the moduli to higher frequencies with increasing temperature, the moduli measured at $\omega = 1$ rad s$^{-1}$ decay with temperature. From this temperature dependence, we characterize the solution as a viscoelastic solid ($G' > G''$) at low temperatures that transitions to a liquid ($G' < G''$) for $T > 8 \: ^{\circ}$C. Notably, the elasticity of this solution changes by over two orders of magnitude for $0 \: ^\circ$C $< T < 30 \: ^\circ$C.

\begin{figure}[tb!]
\centering
\includegraphics[width = 3.25 in]{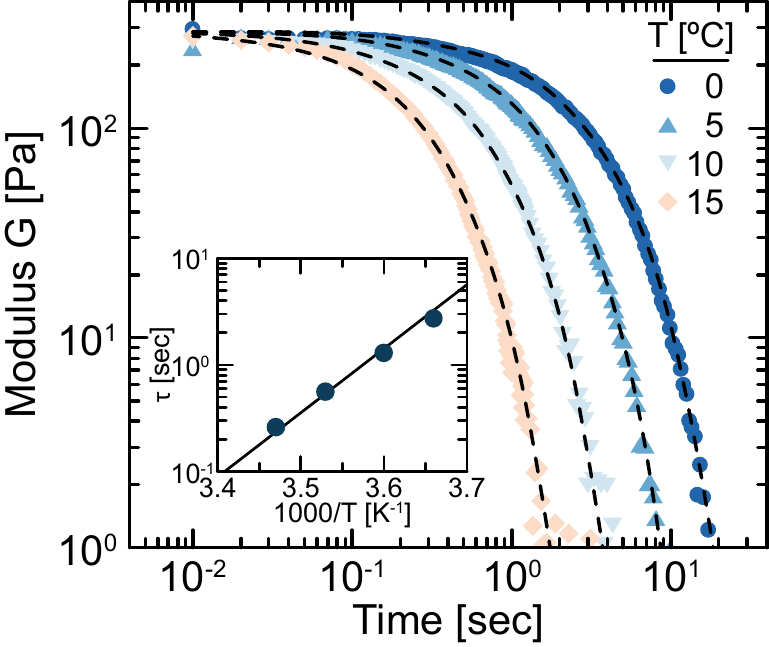}
\caption{\label{fig:WLM_Relaxation} Modulus $G$ after a step strain of $\gamma = 1$\% of the neat WLM solution as a function of time for different temperatures. Dashed curves are KWW fits with $G_0 = 280$ Pa and $\beta = 0.95$. \textit{Inset:} Temperature dependence of relaxation time $\tau$. Solid line indicates Arrhenius behavior with the same $|E_\mathrm{a}|$ as in Fig. \ref{fig:WLM_Tdep}.}
\end{figure}

In addition to the frequency response of the WLM solution, we characterize the stress relaxation of the system (Fig.\ \ref{fig:WLM_Relaxation}). As expected from the Maxwellian frequency response, stress relaxes exponentially in time and is well-fit by a Kohlrausch-Williams-Watts (KWW) expression $G = G_0 \exp\left(-(t/\tau)^\beta\right)$, where $G_0$ is the initial shear modulus, $\tau$ is the system relaxation time, and $\beta$ is a stretching exponent. The stretching exponent $\beta = 0.95$ extracted from these fits is very close to the expected single exponential behavior of a Maxwell fluid. Analogous to the shift in crossover frequency $\omega_\mathrm{c}$ measured in oscillatory rheology, the stress relaxation curves shift to shorter times for higher temperatures. The relaxation times measured from these stress relaxation experiments has the same temperature dependence and activation energy as $\omega_\mathrm{c}$. These results confirm that the WLM solution is a model viscoelastic fluid whose elasticity at a fixed frequency can be controlled by varying temperature.

\subsection{WLM Emulsion}

\begin{figure}[tb!]
\centering
\includegraphics[width = 3.25 in]{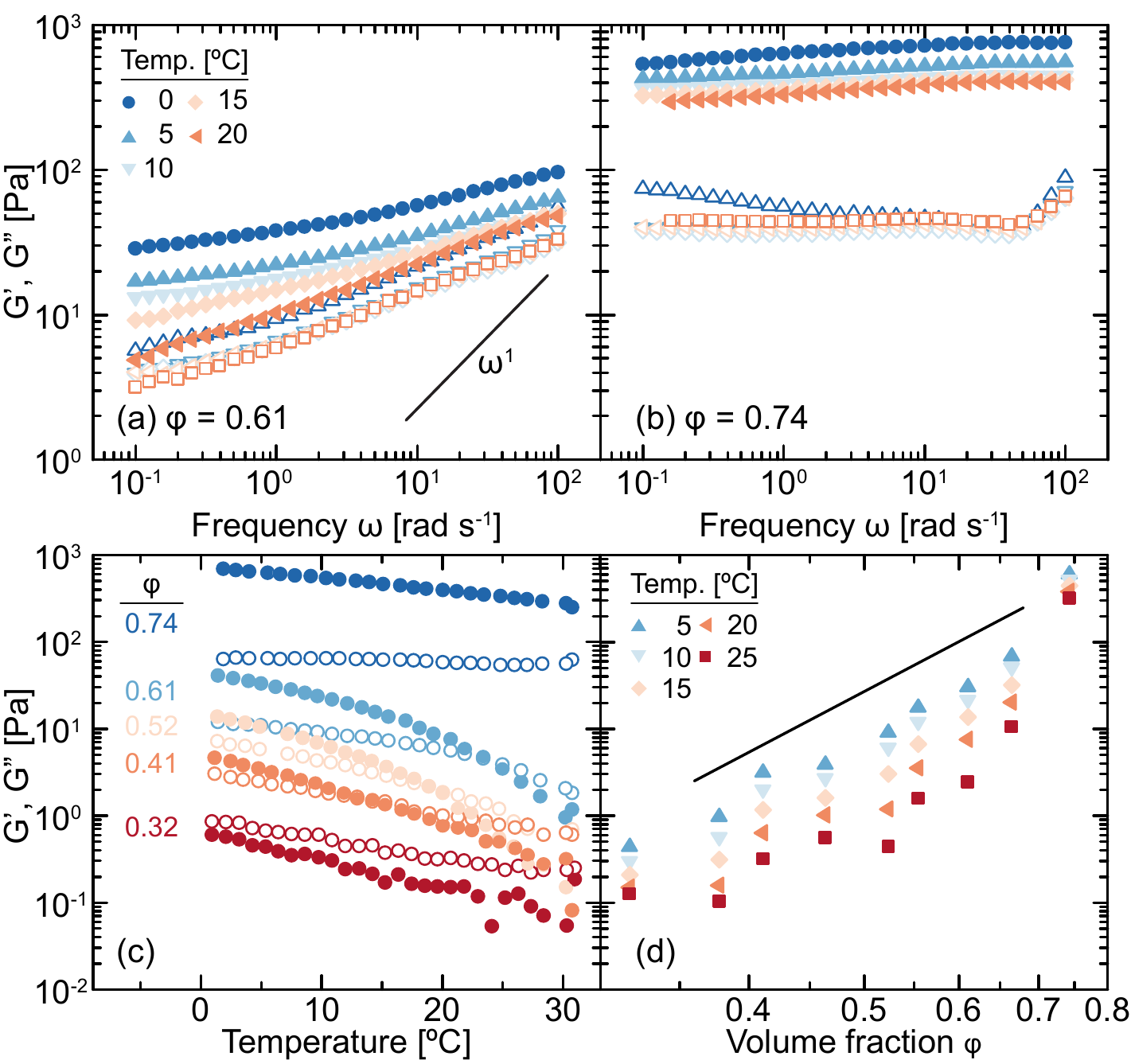}
\caption{\label{fig:TandC} Storage $G'$ (closed) and loss $G''$ (open) moduli of the WLM emulsion as a function of frequency for (a) $\phi = 0.61$ and (b) $\phi = 0.74$ and (c) as a function of temperature at $\omega = 1$ rad s$^{-1}$ for different volume fractions. (d) Storage modulus $G'$ at $\omega = 1$ rad s$^{-1}$ as a function of emulsion volume fraction $\phi$ for various temperatures. Solid line indicates power-law scaling with an exponent of $7.2 \pm 0.5$.}
\end{figure}

The aqueous WLM solution forms a stable emulsion when dispersed in the continuous oil phase. Similar to the characterization of the neat WLM solution, the properties of this WLM emulsion are determined through oscillatory rheology measurements (Fig.\ \ref{fig:TandC}). An emulsion at a volume fraction $\phi = 0.61$ forms a viscoelastic solid with $G' > G''$. Both viscous and elastic moduli increase with increasing frequency and decreasing temperature. More importantly, the low frequency response of G' becomes shallower with decreasing temperature indicating that stress relaxes in these emulsions over longer time scales at low temperatures. At higher volume fractions (\textit{e.g.} $\phi = 0.74$), the emulsion arrests so that the moduli become nearly independent of $\omega$, which is consistent with the behavior of other densely packed materials such as colloidal glasses, dry foams, and other emulsions.\cite{Durian1997,Sollich1997,VanHecke2010,Seth2011,Vlassopoulos2014,Helgeson2016,Bonn2017} 

To assess how these emulsions respond to temperature, we perform a temperature sweep at a specific frequency $\omega = 1$ rad s$^{-1}$ (Fig\ \ref{fig:TandC}(c)). For the non-jammed emulsions, the moduli decrease with increasing temperature so that the emulsions transition from a viscoelastic solid to a viscoelastic liquid at a specific temperature. This transition temperature increases with increasing volume fraction. Furthermore, the storage modulus increases with volume fraction as a power-law $G\sim \phi^\alpha$ with $\alpha = 7.2 \pm 0.5$ for all temperatures and for all volume fractions $\phi < 0.7$. At $\phi = 0.74$, the temperature dependence is effectively suppressed; this is consistent with the emulsion being compressed in this regime. This power-law exponent $\alpha$ is comparable to that measured for suspensions of poly(\textit{N}-isopropylacrylamide) microgel particles.\cite{Koumakis2012} Although the modulus decreases with temperature at all volume fractions, there is no significant change in the power-law exponent with changing temperature despite the elasticity of the neat WLM dropping by over two orders of magnitude. Surprisingly, the change in droplet elasticity does not appear to significantly affect the volume fraction dependence of the modulus of these emulsions. This is in contrast to previously reported studies on dense suspensions of compressible soft objects \cite{Mattsson2009,Lyon2012,Seekell2015, Nigro2017} but consistent with recent studies on non-compressible soft particles.\cite{VanderScheer2017,Philippe2018,Gnan2019} Although the moduli of the emulsified WLM droplets can change with temperature, the droplets always remain incompressible and therefore, the fragility of these suspensions does not change with temperature. In total, these rheological changes indicate that relaxations in this system depend on both the volume fraction and modulus of the dispersed phase.

\begin{figure}[tb!]
\centering
\includegraphics[width = 3.25in]{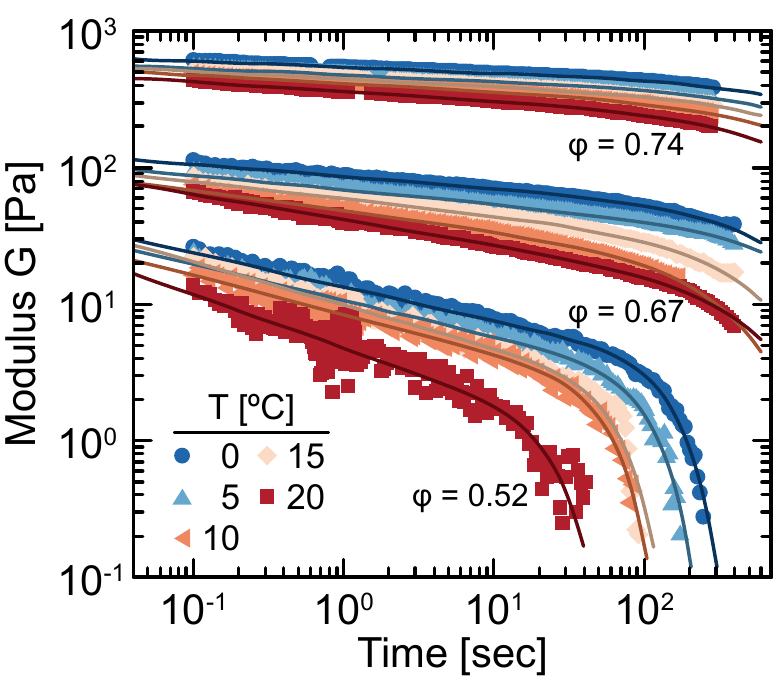}
\caption{\label{fig:EmulRelax} Modulus $G$ after a step strain of $\gamma = 1$\% as a function of time for WLM emulsions at different temperatures and volume fractions. Curves are fits to Eq. \ref{eq:EmulRelax}.}
\end{figure}

For dense systems, stress can relax through a variety of mechanisms. In colloidal suspensions,\cite{Mason1995,Erwin2010,Joshi2014,Poling-Skutvik2019} there is competition between cage-rattling and cage breaking mechanisms that leads to shallow relaxations at short times and quasi-exponential decays at long times.\cite{Segre1996,Watanabe1999,OBrien2000,Yin2008,McKenna2009,Wen2015} In polymer systems,\cite{Rubinstein2003} polymer segments relax subdiffusively through Zimm or Rouse motion before the chain moves diffusively so that the stress response decays as a power-law on short time scales followed by an exponential at long times.\cite{Richter1978,Kapnistos2008,Hou2010} 
We observe a similar complex response in the stress relaxation of the WLM emulsions and choose to model the stress relaxation through an analogous process (Fig.\ \ref{fig:EmulRelax}). We fit the stress response to the product of a power-law decay and a stretched exponential form as shown in eq.\ \ref{eq:EmulRelax}, where $A$ is a pre-factor, $k$ is an exponent characterizing the steepness of the power-law decay, $\tau$ is a relaxation time, and $\beta$ is the stretching exponent.
\begin{equation}
\label{eq:EmulRelax}
G = A t^{-k}\exp\left[-(t/\tau)^\beta\right]
\end{equation}
This expression reasonably fits the complex stress relaxations across all temperatures and volume fractions.

\begin{figure}[tb!]
\centering
\includegraphics[width = 3.25 in]{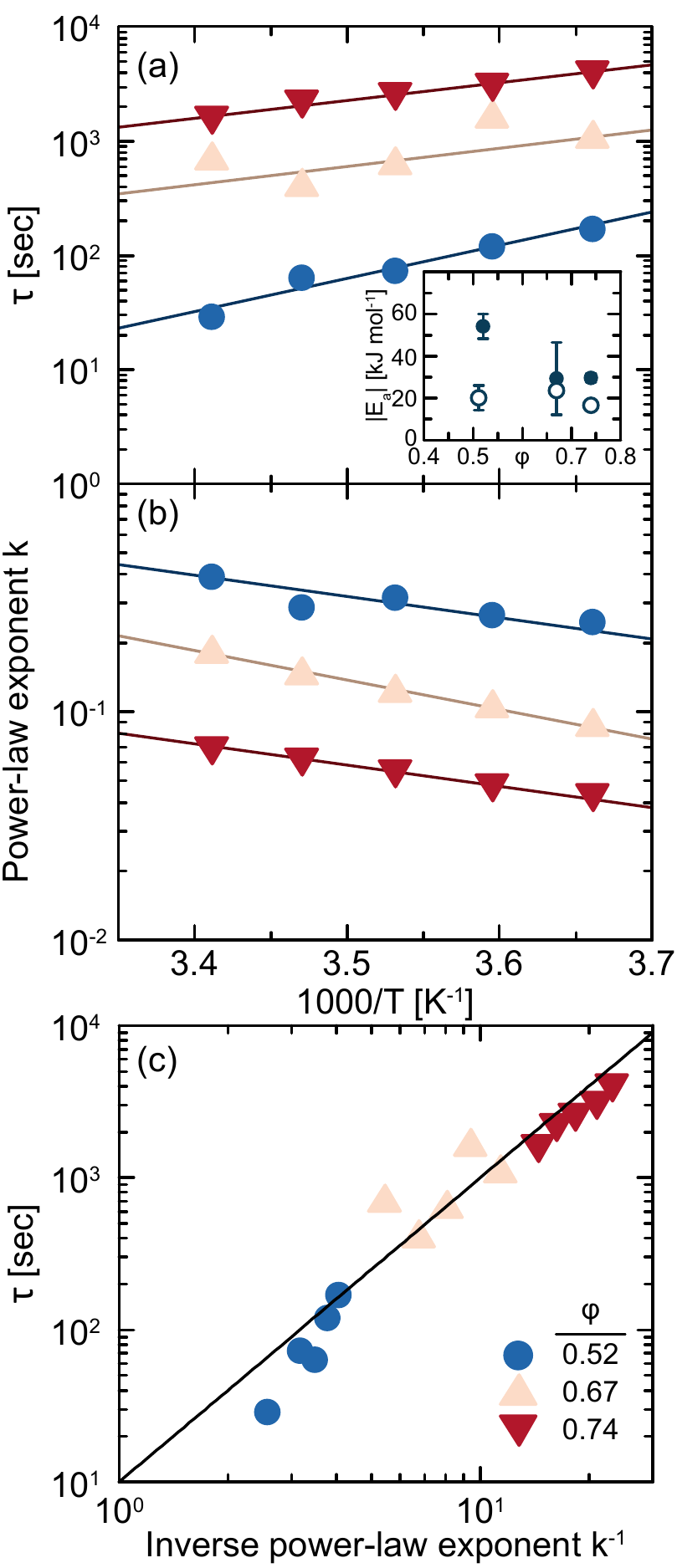}
\caption{\label{fig:RelaxFits} (a) Relaxation time $\tau$ and (b) power-law exponent $k$ as a function of inverse temperature extracted from fits to relaxations in Fig.\ \ref{fig:EmulRelax} using Eq.\ \ref{eq:EmulRelax}. Solid lines are Arrhenius fits. \textit{Inset to (a):} Magnitude of activation energy $|E_\mathrm{a}|$ for $\tau$ (closed) and $k$ (open) as a function of emulsion volume fraction $\phi$. (c) Relaxation time $\tau$ as a function of inverse power-law scaling exponent $k^{-1}$. Solid line shows a guide to the eye.}
\end{figure}

Physically, the relaxation time $\tau$ corresponds to the terminal relaxation of the system. In dense suspensions, this terminal relaxation time scale corresponds to a particle breaking out of the cage formed by neighboring particles and beginning to diffuse.\cite{Segre1996,Hunter2012, Poling-Skutvik2019} As the emulsion volume fraction increases, $\tau$ increases, inidcating that the particle caging becomes stronger so that it is harder for the droplets to move past their neighbors (Fig.\ \ref{fig:RelaxFits}). Additionally, $\tau$ increases with decreasing temperature as the emulsified WLMs stiffen. This increase in droplet relaxation time with decreasing temperature indicates that stiffer, more elastic droplets form stronger cages than softer, more viscous droplets. Similar to the properties of the neat WLMs, this thermal change in $\tau$ follows an Arrhenius expression with $\phi$-dependent activation energies that are lower than that of the neat WLM solution. These Arrhenius dependences indicate that the thermal response of the neat WLM solution leads to a thermal response of the emulsion droplet dynamics. The low values and volume fraction dependence of $E_\mathrm{a}$ indicates, however, that droplet elasticity does not wholly control droplet dynamics. Specifically, increasing the WLM elasticity by over two orders of magnitude (Fig.\ \ref{fig:WLM_Tdep}) slows relaxations by less than one order of magnitude. Intriguingly, as caging becomes stronger with increasing $\phi$, the thermal response of the droplet dynamics becomes weaker. This decrease in $|E_\mathrm{a}|$ with increasing $\phi$ suggests that droplet dynamics depend on a competition between the stress relaxation of the emulsified WLM network and the cage breaking of the emulsion droplets. At very low $\phi$, we would expect emulsion properties to become independent of particle elasticity with a viscosity $\eta = \eta_0 (1+2.5\phi)$ and diffusive droplet dynamics that agree with the Stokes-Einstein expression. At the opposite extreme of very high $\phi$, droplets are completely localized within the neighboring cage so that terminal relaxations are fully suppressed. At the intermediate $\phi$ studied here, however, both droplet dynamics and the WLM elasticity affect emulsion dynamics and stress relaxation mechanisms, leading to the intermediate temperature dependence observed in Fig.\ \ref{fig:RelaxFits}(a).

\begin{figure*}[h!]
\centering
\includegraphics[width = 6.5 in]{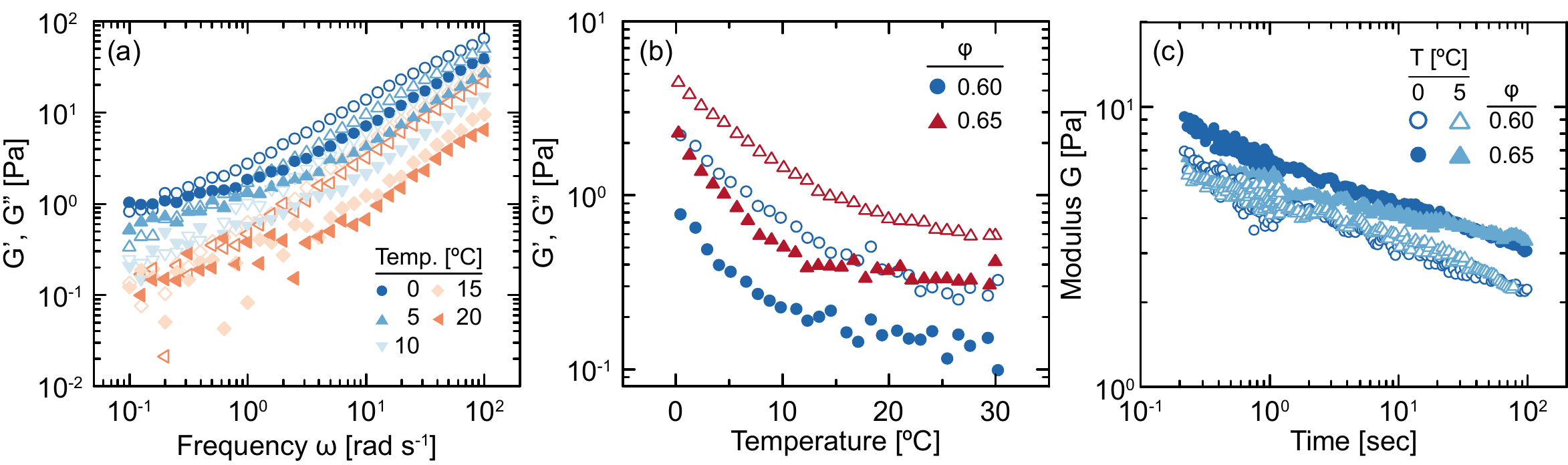}
\caption{\label{fig:WaterEmul} Storage $G'$ (closed) and loss $G''$ (open) moduli as a function of (a) frequency at different temperatures for a water emulsion at $\phi = 0.65$ and of (b) temperature at $\omega = 1$ rad s$^{-1}$ for different volume fractions. (c) Modulus $G$ as a function of time for water emulsions at different volume fractions and temperatures.}
\end{figure*}

On time scales shorter than terminal relaxations, stress in the emulsion partially relaxes through a power-law decay that we speculate arises from particles rattling within cages formed by their nearest neighbors. How particles rattle within such cages depends strongly on volume fraction, which controls the size of the cage,\cite{Hunter2012} as well as on the particle softness, which controls the steepness of the short range inter-particle interaction.\cite{Zaccarelli2008, Ikeda2011, Pastore2014} As volume fraction increases, cage size decreases so that rattling is suppressed. Subsequently, stress relaxes more slowly with time at higher volume fractions, consistent with the lower power-law exponents $k$ at higher $\phi$ (Fig.\ \ref{fig:RelaxFits}(b)). Similarly, softer particles (i.e. those interacting via less steep inter-particle potentials) rattle more in their cages and can relax stress more quickly than hard particles. Due to the thermoresponsive nature of the WLMs in the emulsion droplets, the droplets lose elasticity and become softer with increasing temperature. The measured power-law exponent $k$ subsequently increases with increasing temperature. This increase follows an Arrhenius dependence with activation energies are largely independent of volume fraction, $\phi$. The volume fraction dependence of $k$ observed here is in contrast to the insensitivity to volume fraction seen in power-law stress relaxation in dilute colloidal gels, where local structure is independent of particle volume fraction.\cite{Negi_stress2009}. Moreover, $k$ is inversely correlated with $\tau$ (Fig.\ \ref{fig:RelaxFits}(c)), indicating that faster stress relaxation on short time scales leads to faster terminal relaxation. The collapse in the stress relaxation behavior identified by the correlation between $\tau$ and $k$ indicates that emulsion droplet elasticity and volume fraction are independent parameters that control the dynamics of dense emulsions. Accordingly, an increase in droplet stiffness slows dynamics in the same fashion as an increase in the emulsion volume fraction. Although it makes intuitive sense that slower relaxations on short time scales lead to slower terminal relaxations, the physical origin of a relationship between $\tau$ and $k$ remains elusive. We hypothesize that multiple parameters control the coupling between short and long time relaxations, including droplet size, the viscosity of the continuous phase, and droplet surface chemistries. Changes in these parameters would likely shift the collapse to different values of $\tau$ and $k$ and perhaps affect the functional relationship between them.

\subsection{Water emulsion}

As a control, we examine the behavior of a dense suspension of emulsified water droplets without WLMs prepared using the same methods (Fig.\ \ref{fig:WaterEmul}). First, at similar volume fractions, the water emulsion forms a viscoelastic fluid ($G' < G''$) with much lower elasticity than the WLM emulsion. The frequency response of the water emulsion is similar to that of the WLM emulsion at higher temperatures (\textit{i.e.} when the WLM network is viscous) and dissimilar from the response of the WLM emulsion at low $T$ (\textit{i.e.} when the WLM network is elastic). Second, the moduli of the water emulsion decreases slightly with increasing temperatures from $0 \:^\circ\mathrm{C} < T < 10 \:^{\circ}\mathrm{C}$ but plateaus at higher temperatures without any further temperature dependence. This behavior strongly contrasts with the temperature dependence of the moduli for the WLM emulsion which decrease by over a magnitude and never plateau over a similar temperature range. Third, the water emulsions relax stress through power-law mechanisms similar to that of the emulsified WLMs as the water droplets rattle in their local cages. Similar to the emulsified WLM samples, the power-law relaxations are shallower at higher volume fractions. By contrast, however, the stress relaxations of the water emulsion do not vary with temperature. Thus, the strong thermoresponsive behavior of the emulsified WLMs is unique to the WLM system and arises from changes to the internal elasticity of the droplets. 

\section{Conclusion}
We demonstrate that wormlike micelles can be successfully emulsified in a continuous oil phase using orthogonal surfactants to produce emulsions with thermoresponsive characteristics. In this model system, the droplets are elastic at low temperatures and viscous at high temperatures allowing us to probe the effect of droplet elasticity on emulsion properties independent of droplet compressibility. The emulsion modulus increases with increasing droplet volume fraction as a power-law with a temperature-independent exponent, indicating that the emulsion fragility does not depend on droplet elasticity. By contrast, the droplet dynamics are strongly dependent on the droplet elasticity. As the droplets become less elastic with increasing temperature, the droplets escape their local cages faster, leading to faster terminal relaxations. We observe a coupling between the two modes of stress relaxation, with both volume fraction and temperature independently affecting the power law stress relaxations on short time scales and exponential modes on long time scales. Our findings indicate that the viscoelasticity of emulsion droplets significantly alters droplet dynamics and how stress relaxes through dense emulsions.

\section*{Conflicts of interest}
There are no conflicts to declare.

\section*{Acknowledgments}
This work was supported by a grant from NSF (CBET 1066904). C. O. and R. P-S. acknowledge additional support from NSF (DMR 1947707 and CMMI 1548571)

\newpage

\bibliography{biblio} 

\newpage
\section*{For table of contents only:}
\begin{centering}
\includegraphics[width = 8cm]{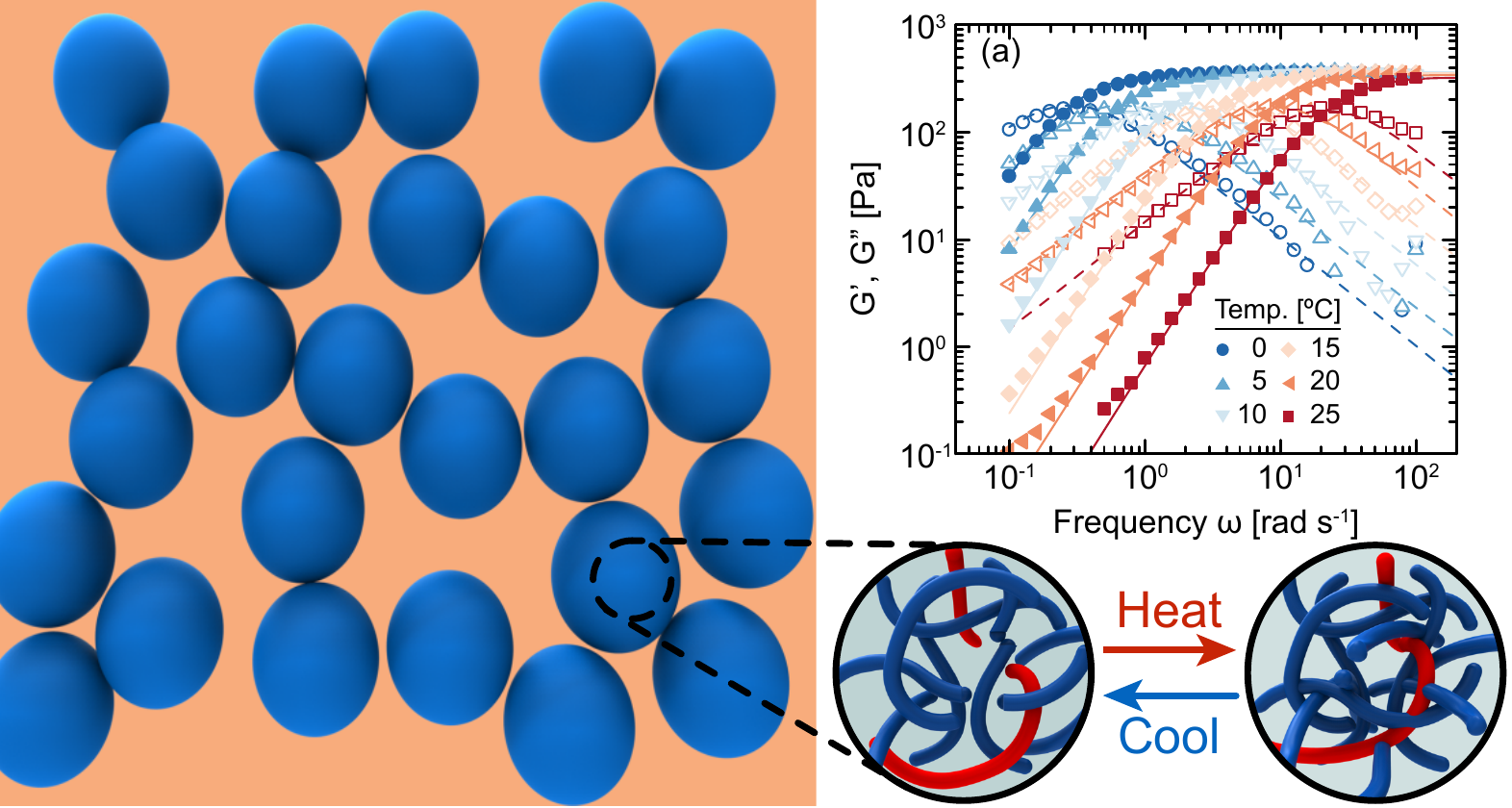}
\end{centering}

\end{document}